\documentclass[10pt]{article}
\usepackage[dvips]{graphicx}
\usepackage{mathrsfs}
\usepackage{eucal}
\usepackage{amsmath,amsthm,amssymb}
\usepackage{wrapfig}
\usepackage[dvips]{color}
\usepackage{framed}
\usepackage{tabularx}
\usepackage{threeparttable}
\definecolor{shadecolor}{gray}{0.80}
\DeclareMathVersion{chem}
\SetSymbolFont{letters}{chem}{OT1}{cmr}{m}{n}

\begin{document}

\renewcommand{\figurename}{\small{Fig.}~}
\renewcommand{\labelitemi}{}
\renewcommand{\thefootnote}{$\dagger$\arabic{footnote}}
\renewcommand{\footnoterule}{%
  \vspace{2pt}
 \flushleft\rule{6.154cm}{0.4pt}
  \vspace{4pt}
  }
\pagestyle{plain}

\begin{flushright}
\textit{Concentration Invariance of Cyclic Species}
\end{flushright}
\vspace{1mm}

\begin{center}
\setlength{\baselineskip}{25pt}{\LARGE\textbf{Concentration Invariance of Cyclic Species}}
\end{center}
\vspace{0mm}

\vspace*{0mm}
\begin{center}
\large{Kazumi Suematsu} \vspace*{2mm}\\
\normalsize{\setlength{\baselineskip}{12pt} 
Institute of Mathematical Science\\
Ohkadai 2-31-9, Yokkaichi, Mie 512-1216, JAPAN\\
E-Mail: suematsu@m3.cty-net.ne.jp,  Tel/Fax: +81 (0) 593 26 8052}\\[8mm]
\end{center}

\hrule
\vspace{0mm}
\begin{flushleft}
\textbf{\large Abstract}
\end{flushleft}
Concentration invariance of cyclic species in the irreversible polymerization is examined. The simulation shows that the invariance theorem holds in good approximation for the irreversible process also. The physical soundness of the underlying simulation equations is confirmed through comparison with the recent experimental observations carried out by the Nagoya university group.\\[-3mm]
\begin{flushleft}
\textbf{Key Words}:
\normalsize{Irreversible Polymerization/ Distribution of Cyclic Species/ Concentration Invariance}\\[3mm]
\end{flushleft}
\hrule
\vspace{3mm}
\setlength{\baselineskip}{13pt}
\section{Introduction}
It is well established that the relative frequency of ring formation to intermolecular reaction increases with dilution. It was pointed out, on the other hand, that the ring concentration is independent of the monomer concentration\cite{Kazumi}, which has been proven rigorously for the reversible difunctional system. In this paper we investigate the irreversible polymerization\cite{Yuan} to examine whether the principle of the concentration invariance of cyclic species holds for the irreversible polymerization also. Prior to the examination, we assess the physical soundness of the underlying simulation equations through comparison with the recent experimental observation for the AA-type telechelic polystyrene carried out by the Nagoya university group \cite{Kushida}.

\section{Theoretical}
We first assess the soundness of the basic equations under the assumption of the equal reactivity of functional units. To deal with the irreversible process, we must introduce another assumption that the system obeys the quasi-static reaction so that the reaction mixture is approximately in equilibrium and the complete mixing of reaction mixture is assured in all reaction stages. 

Let $M_{0}$ be the initial monomer number and $V$ the system volume. The birth-death equation for the AA-type bifunctional polymerization can be expressed in the form:
\begin{align}
\delta N_{x}/\delta u&=\frac{2\displaystyle\sum\nolimits_{j=1}^{x-1}N_{j}\,N_{x-j}}{2(M_{0}-u)^{2}+V\displaystyle\sum\nolimits_{k=1}^{\infty}\varphi_{k}N_{k}}-\frac{4N_{x}(M_{0}-u)+V\varphi_{x}N_{x}}{2(M_{0}-u)^{2}+V\displaystyle\sum\nolimits_{k=1}^{\infty}\varphi_{k}N_{k}}\\
\delta N_{R_{x}}/\delta u&=\frac{V\varphi_{x}N_{x}}{2(M_{0}-u)^{2}+V\displaystyle\sum\nolimits_{k=1}^{\infty}\varphi_{k}N_{k}},
\end{align}
where $N_{x}$ and $N_{R_{x}}$ denote the numbers of chain $x$-mers and cyclic $x$-mers respectively with $x=1,2,\cdots$ being the number of repeating units, $\varphi_{x}$ the relative cyclization frequency of an $x$-meric chain, and $u$ represents the number of unit reactions ($u=0,1,2,\cdots$). For bifunctional reactions, $u$ is equal to the number of bonds. The quantity $\varphi_{x}$ is defined by $\mathscr{P}_{x}/v$\cite{Kazumi} where $\mathscr{P}_{x}$ is the probability that one end of the $x$-meric chain enters a small volume $v$ with the radius $\ell_{s}$ around the another end. For the Gaussian chain, $\varphi_{x}$ has the form:
\begin{equation}
\varphi_x=\big(d/2\pi^{d/2}\ell_{s}^{\hspace{0.3mm}d}N_{A}\big)\displaystyle\int_{0}^{d/2\nu_{\hspace{-0.3mm}x}}\hspace{-2mm}t^{\frac{d}{2}-1}e^{-t}dt,
\end{equation}
where \textit{d} is the space dimension, $N_{A}$ the Avogadro number and $\nu_{x}$ the quantity defined by
\begin{equation}
\nu_{x}=\langle r_{x}^{2}\rangle/\ell_{\hspace{-0.3mm}s}^{\hspace{0.3mm}2},
\end{equation}
where $\langle r_{x}^{2}\rangle$ is the mean square end-to-end distance of the $x$-mer chain. In terms of the excluded volume effect, it becomes: $\langle r_{x}^{2}\rangle=\alpha^{2}\langle r_{x}^{2}\rangle_{\varTheta}$, with $\alpha$ being the expansion factor and a function of $x$. The subscript $\varTheta$ denotes the unperturbed state in which the excluded volume effect is expected to vanish, which is realizable in poor solvents, or in concentrated solutions. Then
\begin{equation}
\nu_{x}=\alpha^{2}\langle r_{x}^{2}\rangle_{\varTheta}/\ell_{\hspace{-0.3mm}s}^{\hspace{0.3mm}2}=\alpha^{2}C_{F}\hspace{0.2mm}\xi_{e}\hspace{0.3mm}x\nonumber\eqno (4')
\end{equation}
where $C_{F}$ signifies the Flory characteristic ratio, and $\xi_{e}$ the effective bond number within a repeating unit defined by $\xi_{e}=\frac{1}{\ell_{s}^{2}}\sum_{i}\ell_{i}^{2}$. Note that $C_{F}$ is not a constant, but an increasing function of $x$, approaching a constant as $\xi_{e}x\rightarrow\infty$.

Contrary to the ideal system without rings, there is no explicit closed solution for the equations (1) and (2); so $N_{x}$ or $N_{R_{x}}$ can not be expressed uniquely in terms of $u$. Fortunately, the equations can be solved numerically \cite{Yuan}. Chemical reactions are by nature discrete, because the number of events is enumerable in principle. Thus the above equations (1) and (2) are recursive in essence. To find numerical solutions, it is convenient to sum up these equations from $u=0$ to $z-1$ $(0\le z\le M_{0})$ to obtain
\begin{align}
N_{x}(z)&=\sum_{u=0}^{z-1}\left\{\frac{2\displaystyle\sum\nolimits_{j=1}^{x-1}N_{j}(u)\,N_{x-j}(u)}{2(M_{0}-u)^{2}+V\displaystyle\sum\nolimits_{k=1}^{\infty}\varphi_{k}N_{k}(u)}-\frac{4N_{x}(u)(M_{0}-u)+V\varphi_{x}N_{x}(u)}{2(M_{0}-u)^{2}+V\displaystyle\sum\nolimits_{k=1}^{\infty}\varphi_{k}N_{k}(u)}\right\}\\
N_{R_{x}}(z)&=\sum_{u=0}^{z-1}\left\{\frac{V\varphi_{x}N_{x}(u)}{2(M_{0}-u)^{2}+V\displaystyle\sum\nolimits_{k=1}^{\infty}\varphi_{k}N_{k}(u)}\right\},
\end{align}
with the boundary conditions:
\begin{equation}
N_{x}(0)=
\begin{cases}
\hspace{1mm}0 & \text{if}\hspace{2mm} x\neq 1\\
M_{0} & \text{if}\hspace{2mm} x=1, \nonumber
\end{cases}
\hspace{7mm} N_{R_{x}}(0)=0\hspace{3mm} \text{for all $x$'s}.
\end{equation}

Let us perform the calculation of the above equations modeling the polymerization of the telechelic polystyrene monomer (t-PSt: M$_\text{w}=4.6\times 10^{4}$, $\text{M}_{\text{w}}/\text{M}_{n}=1.02$) at $25^{\circ}$C in tetrahydrofuran (THF) as a good solvent, with the help of the parameters shown in Table 1 \cite{Kushida}.

\begin{center}
\begin{threeparttable}
\caption{Model polymerization of the telechelic polystyrene (t-PSt) in THF}
\vspace{-2mm}
\begin{tabular}{l c c}\hline\\[-4mm]
parameter & \hspace{3mm}symbol & \hspace{3mm}value \\[1mm]
\hline\\[-2mm]
molecular weight & \hspace{3mm}M$_{\text{w}}$& \hspace{3mm} $4.6\times 10^{4}$\\[1mm]
characteristic ratio & \hspace{3mm}$C_{F}$ & \hspace{3mm} 10\\[1mm]
effective bond number & \hspace{3mm}$\xi_{e}$ & \hspace{3mm} 867\\[1mm]
standard bond length $\left(\text{\AA}\right)$& \hspace{3mm}$\ell_{s}$ & \hspace{3mm} 1.55\\[1mm]
expansion factor & \hspace{3mm}$\alpha$ & \hspace{3mm} $1.43\times x^{0.06}$\\[1mm]
monomer concentration ($\,mol/l$) & \hspace{3mm}$C_{0}$ & \hspace{3mm} $2\times10^{-5}\,\tnote{a}$\\[2mm]
\hline\\[-7mm]
\end{tabular}
   \vspace*{2mm}
   \begin{tablenotes}
     \item a. which corresponds to $\phi\cong0.001$.
   \end{tablenotes}
  \end{threeparttable}
\end{center}
\vspace*{3mm}

\noindent This system corresponds to $\phi\cong 0.001$\,(volume fraction of polymer), close to the dilution limit. Thus the classic excluded volume argument applies. Consulting experimental reports\cite{Venkataswamy, Fetters} on coil expansion, we find that $\alpha$ (expansion factor) $=1.43\times x^{0.06}$\,\footnote{\, estimated from the empirical equation at 30$^\circ$C by Venkataswamy and coworkers\cite{Venkataswamy}.}. The mean square end-to-end distance then becomes $\langle r_{x}^{2}\rangle=\alpha^{2}\langle r_{x}^{2}\rangle_{\varTheta}=1.43^{2}C_{F}\hspace{0.2mm}\xi_{e}\hspace{0.3mm}x^{1.12}\hspace{0.3mm}\ell^{2}$. Since $\xi_{e}$ is sufficiently large\cite{Flory}, we may approximate $C_{F}\simeq$ constant for all $x$'s.

The simulation was performed to calculate the weight fractions of cyclic, $w_{R}=\sum_{x}xN_{R_{x}}/M_{0}$, and acyclic species, $w_{C}=\sum_{x}xN_{x}/M_{0}$, at $D=0.84$ (Fig. 1), where $D$ is the extent of the advancement of reaction defined by $D=u/M_{0}$. The results are summarized in Fig. 1 and Table 2.

\vspace{3mm}
\begin{figure}[h]
\begin{center}
\includegraphics[width=7.0cm]{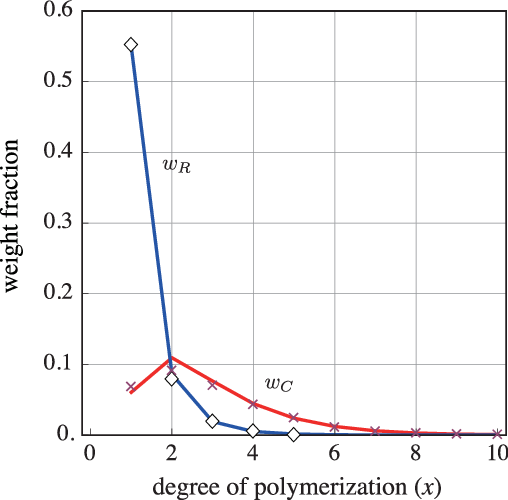}
\begin{center}
\vspace{-5mm}
\parbox[t]{93mm}{\caption{Weight fractions of cyclic and acyclic species in the polymerization of the telechelic polystyrene monomer ($\text{M}_{\text{w}}=46000$). Diamonds ($\diamondsuit$) and crosses ($\times$): experimental weight fractions of cyclic and acyclic species, respectively, by Kushida and coworkers \protect\cite{Kushida}. Solid lines ($-$): theoretical lines at $D=0.84$ by eqs. (5) and (6).}}
\end{center}
\end{center}
\end{figure}
\vspace*{-4mm}

\begin{center}
\begin{threeparttable}[h]
\caption{Comparison of the theory and the experiments}
\begin{tabular}{c c c c c c c}\hline\\[-3mm]
\multicolumn{7}{c}{Weight fraction of acyclic and cyclic species at $D=0.84$}\\[1mm]
\hline\\[-2mm]
\multicolumn{1}{c}{} && \multicolumn{2}{c}{Theoretical} && \multicolumn{2}{c}{Observed\,\tnote{a}}\\[1mm]
\cline{3-4}\cline{6-7}\\[-3mm]
\multicolumn{1}{c}{$n$-mer} && chains & rings && chains & rings \\[1mm]
\hline\\[-2mm]
1 && \hspace{3mm}0.06051 & \hspace{3mm} 0.55009 && \hspace{3mm}0.068318 & \hspace{3mm} 0.55188\\[1mm]
2 && \hspace{3mm}0.10935 & \hspace{3mm} 0.08581 && \hspace{3mm}0.091091 & \hspace{3mm} 0.07884\\[1mm]
3 && \hspace{3mm}0.07633 & \hspace{3mm} 0.01897 && \hspace{3mm}0.070070 & \hspace{3mm} 0.01905\\[1mm]
4 && \hspace{3mm}0.04445 & \hspace{3mm} 0.00499 && \hspace{3mm}0.042918 & \hspace{3mm} 0.00521\\[1mm]
5 && \hspace{3mm}0.02381 & \hspace{3mm} 0.00145 && \hspace{3mm}0.024087 & \hspace{3mm} 0.00000\\[1mm]
6 && \hspace{3mm}0.01214 & \hspace{3mm} 0.00045 && \hspace{3mm}0.010948 & \hspace{3mm} 0.00000\\[2mm]
\hline\\[-7mm]
\end{tabular}
   \vspace*{2mm}
   \begin{tablenotes}
     \item a. Experimental $D$ value is unavailable\,\protect\cite{Kushida}.
   \end{tablenotes}
  \end{threeparttable}
\end{center}
\vspace*{4mm}

\noindent In Fig. 1, solid lines represent the theoretical curves for cyclic (blue solid line) and acyclic species (red solid line) based on eqs. (5) and (6) ($D=0.84$), and diamonds ($\diamondsuit$) and crosses ($\times$) are corresponding experimental observations by Nagoya University group \cite{Kushida} under the same reaction condition ($D=0.84$ is assumed). Agreement between the theory and the experiments is very excellent, in support of the physical soundness of the kinetic formulas (5) and (6).

\newpage
\begin{wrapfigure}[20]{r}{7.3cm}
\vspace*{-3mm}
\begin{center}
\includegraphics[width=7.2cm]{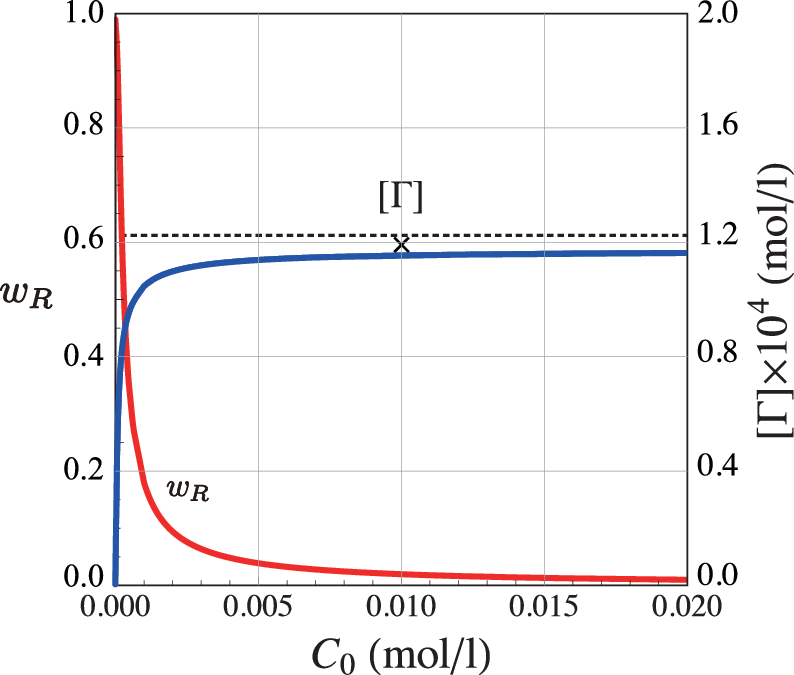}
\end{center}
\vspace{-3mm}
\caption{Total concentration of cyclic species and the weight fraction as against initial monomer (t-PSt) concentration at $D=1$. $[\Gamma]$ and $w_{R}$ were calculated according to eqs. (5) and (6). The dotted line ($\cdots$) shows eq. (8). The cross point ($\times$) is a result for M$_{0}=1000$.}
\end{wrapfigure}

Based on the above results, let us investigate variations of the concentration of cyclic species and the weight fraction as a function of the initial monomer concentration, $C_{0}=M_{0}/V$. The simulation of eqs. (5) and (6) was carried out assuming the $\varTheta$ condition\footnote{\,Otherwise, the cyclization probability must change with concentration because of the excluded volume effects. }, namely we put $\alpha=1$, while holding the other parameters to the same values. 

The results are shown in Fig. 2 for the total ring concentration, $[\Gamma]=\sum_{x} N_{R_{x}}/V$, and the weight fraction, $w_{R}=\sum_{x}xN_{R_{x}}/M_{0}$, at $D\rightarrow 1$; the upper bound of $C_{0}\,\,(\cong 0.02)$ corresponds to the non-solvent state. It is seen that $[\Gamma]$ has the $\Gamma$-letter curve which was already observed in the equilibrium polymerization (see Appendix); i.e., there is a plateau zone in high concentration where $[\Gamma]$ is approximately constant over wide concentration range, whereas in the low concentration of $C_{0}\lesssim 10^{-3}\,\, mol/l$, $[\Gamma]$ decreases rapidly with increasing dilution along with the sudden rise of the weight fraction $w_{R}$ of rings. The sudden rise of $w_{R}$ in the low concentration indicates that the ring formation is going to overwhelm the intermolecular reaction. Below that region the production of large molecules is no longer possible since the molecular growth can occur only through intermolecular linking. Note that eqs. (5) and (6) have the limiting solution of $C_{0}\rightarrow\infty$:

\begin{equation}
[\Gamma]_{C_{0}\rightarrow\infty}=\displaystyle\sum_{x=0}^{\infty}\varphi_{x}\frac{D^{x}}{2x}
\end{equation}
As $D\rightarrow 1$, this leads to

\begin{equation}
[\Gamma]_{C_{0}\rightarrow\infty}\rightarrow\displaystyle\sum_{x=0}^{\infty}\varphi_{x}\frac{1}{2x}
\end{equation}

The dotted line in Fig. 2 represents this eq. (8). It is seen that there is slight numerical difference in the plateau zone between the simulation curve (blue solid line) and eq. (8) (dotted line). There is also observed some difference in the quality of the \textit{concentration invariance} of $[\Gamma]$ between the irreversible case in Fig. 2 and the reversible case in Fig. 4 (see Appendix).  The differences is due to the small sample space ($M_{0}=100$) employed in this simulation; \,the discrepancy is much improved for $M_{0}=1000$ (see the cross point ($\times$) in Fig. 2). The essential feature is, however, quite alike between Fig. 2 and Fig. 4. For both the cases, $[\Gamma]$ is almost constant in the high concentration range. From these results we may conclude that the invariance principle holds also for the irreversible polymerization\cite{Kazumi}.

An important question is whether or not, one can extend the principle of the concentration invariance of cyclic species to a more general case of $f\geq3$ \cite{Kazumi}. At present no rigorous answer is available. From the chemical dynamics point of view, however, the two systems (linear and branching) have the identical feature that the molecular growth occurs through the competition between the intermolecular reaction of \textit{linear molecules} and the cyclization. In branching systems the \textit{linear molecules} are embedded in branched molecules (see Fig. 3).  In this sense, no difference exists between the linear system and the branching system: one may regard every reaction in a branching system as a reaction of a chain molecule (cyclization) or one between chain molecules (intermolecular reaction). Indeed we can describe the ring-forming process of the branching system by the formula exactly identical with that of the linear system:

\begin{equation}
\delta\sum_{x=1}^{\infty} N_{R_{x}}=\frac{\displaystyle\sum\nolimits_{x}\left(v_{R_{x}}/v_{L}\right)}{1+\displaystyle\sum\nolimits_{x}\left(v_{R_{x}}/v_{L}\right)}\,\,\delta u.
\end{equation}

\begin{figure}[h]
\vspace*{-0mm}
\begin{center}
\includegraphics[width=5.3cm]{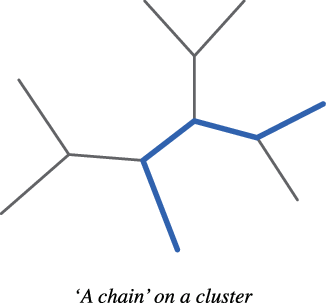}
\end{center}
\vspace{-7mm}
\begin{center}
\parbox[t]{120mm}{\caption{Representation of a chain molecule embedded in a branched cluster.}}
\end{center}
\end{figure}

\noindent For a large $C_{0}$, it follows that $\sum\nolimits_{x}\left(v_{R_{x}}/v_{L}\right)\ll 1$, and the above equation gives the known limiting solution, for instance, $[\Gamma]=\sum_{x}N_{R_{x}}/V=\sum_{x}\varphi_{x}\left[(f-1)D\right]^{x}/2x$ for the R$-$A$_{f}$ model \cite{Kazumi}, which is independent of $C_{0}$. This convinces us the applicability of the concentration invariance of the cyclic production to the branching process.

\section{Possibility of Experimental Verification of the Concentration Invariance}
Let $\Omega(u)$ denote the total number of molecules at $u$ in a given branching system and $\Omega_{0}$ the corresponding quanatity in the ideal system with no rings. Following the known relation, we write
\begin{equation}
\Omega(u)-\Omega_{0}(u)=\Gamma(u),
\end{equation}
while
\begin{gather}
X_{n}=M_{0}/\Omega(u)\\
\Omega_{0}(u)=M_{0}(1-\tfrac{1}{2}fD)
\end{gather}
where $X_{n}$ is the number average molecular weight. Eq. (10) is a universal relation valid for all bond forming processes, if we are in the interval $0\le D\le D_{c}$. Substituting eqs. (11) and (12) into eq. (10) and using the relation $C_{0}=M_{0}/V$ and $[\Gamma]=\Gamma/V$, one has the equality:
\begin{equation}
[\Gamma]=C_{0}\left(\frac{1}{X_{n}}-1+\tfrac{1}{2}fD\right).
\end{equation}
The suggestion of eq. (13) is significant.  It states that if $X_{n}$ can be measured experimentally as a function of $D$ and $C_{0}$, one can estimate $[\Gamma]$ as a function of $C_{0}$. This will provide a possibility that one can test experimentally the assumption of the concentration invariance of $[\Gamma]$ for general branching processes of $f\ge 3$. This possibility was first suggested by Faliagas\cite{Faliagas}. Eq. (13) can be conveniently recast in the form:
\begin{equation}
X_{n}=\frac{1}{[\Gamma]/C_{0}+1-\frac{1}{2}fD}.
\end{equation}
From Fig. 2, the quantity $[\Gamma]/C_{0}$ is of the order $\approx 10^{-2}$. Hence the effect of the ring formation on $X_{n}$ is experimentally  determinable. 
\section{Conclusion}
The present simulation showed that the concentration invariance holds in good approximation for the irreversible difunctional polymerization. There are convincing reasons that we may consider that this principle is applicable to the branching process also, i.e., (i) the limiting solution is independent of $C_{0}$, (ii) no inconsistency has been observed to date for the theory of gelation constructed on the basis of this invariance principle\cite{Kazumi}. There is a possibility to prove experimentally the concentration invariance of cyclic species through the relationship of eq. (14). 

\section*{Appendix}
It will be useful to compare the simulation result in the text with the equilibrium solution\cite{Kazumi}. Suppose an imaginary case in which the above-mentioned t-PSt undergoes the equilibrium polymerization. The corresponding equilibrium solution is of the form:
\begin{equation}
[\Gamma]=\sum_{x=0}^{\infty}\frac{\varphi_{x}}{2x}\left(\frac{D-w_{R}}{1-w_{R}}\right)^{x},
\end{equation}
so that
\begin{equation}
C_{0}=\frac{1}{w_{R}}\sum_{x=0}^{\infty}\frac{\varphi_{x}}{2}\left(\frac{D-w_{R}}{1-w_{R}}\right)^{x}.
\end{equation}
We show a special solution of $D\rightarrow 1$ in Fig. 4.

\begin{figure}[h]
\vspace{2mm}
\begin{center}
\includegraphics[width=15cm]{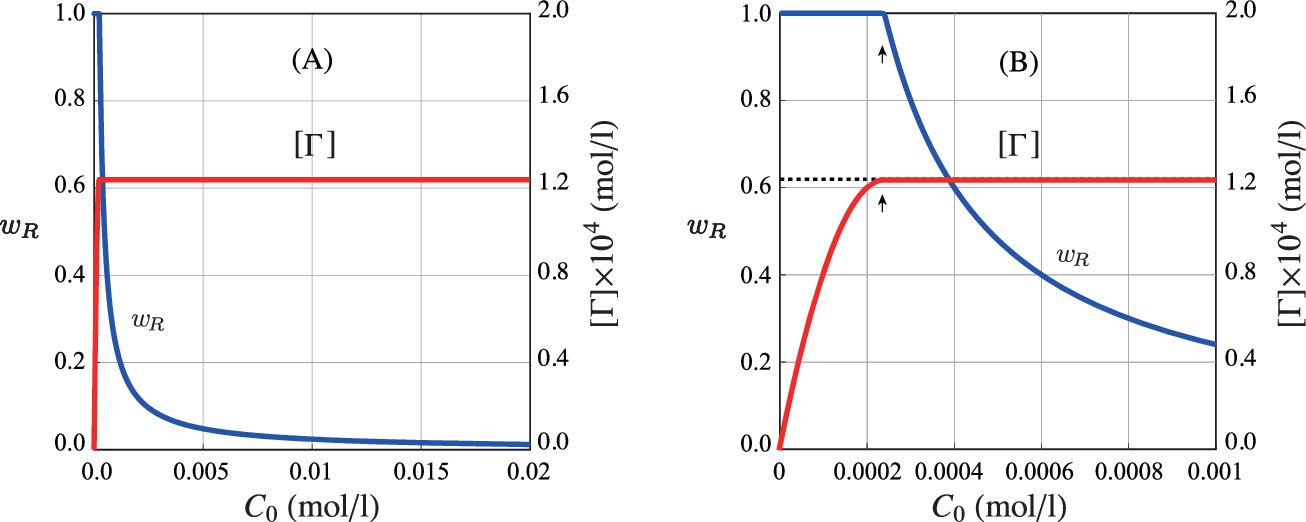}
\parbox[t]{160mm}{\caption{Total concentration $[\Gamma]$ and weight fraction $w_{R}$ of cyclic species in hypothetical equilibrium polymerization as against initial monomer (t-PSt) concentration $C_{0}$. Fig. 4-(B) is a magnification of Fig. 4-(A). The arrows $(\uparrow)$ in Fig. 4-B indicate the critical concentration, $C_{0}^{*}$, below which the concentration invariance of cyclic species breaks down.}}
\end{center}
\end{figure}

As one can see, there is a clear-cut critical point $(\uparrow)$, $C_{0}^{*}$, below which the concentration invariane of $[\Gamma]$ breaks down. From eq. (16), this critical point corresponds to $C_{0}^{*}=\sum_{x=0}^{\infty}\varphi_{x}/{2}.$ Below $C_{0}^{*}$, the weight fraction of rings, $w_{R}$, is held unity, while $[\Gamma]$ begins to decrease. With increasing dilution, $[\Gamma]$ decreases, first slowly, then rapidly, and finally as $[\Gamma]=C_{0} $, showing that the monomeric ring occurs more abundantly with dilution and exclusively in the limit of $C_{0}\rightarrow 0$.\\[-2mm]


\end{document}